\documentclass[conference]{IEEEtran}
\IEEEoverridecommandlockouts
\usepackage[sort&compress, numbers]{natbib}
\usepackage{amsmath,amssymb,amsfonts}
\usepackage{graphicx}
\usepackage{listings}
\usepackage{textcomp}
\usepackage{xcolor}
\usepackage{outline}
\usepackage{multirow}
\usepackage[hyphens]{url}
\usepackage[colorlinks]{hyperref}

\usepackage{enumitem}

\newcommand\sparbox[2]{\parbox[top][#1][c]{6cm}{\strut#2\strut}}

\usepackage{array}
\usepackage{xspace}
\usepackage{todonotes}
\newcolumntype{P}[1]{>{\centering\arraybackslash}p{#1}}

\definecolor{codegreen}{rgb}{0,0.6,0}
\definecolor{codegray}{rgb}{0.5,0.5,0.5}
\definecolor{codepurple}{rgb}{0.58,0,0.82}
\definecolor{backcolour}{rgb}{0.95,0.95,0.92}

\lstdefinestyle{pythonstype}{
    backgroundcolor=\color{backcolour},   
    commentstyle=\color{codegreen},
    keywordstyle=\color{magenta},
    numberstyle=\tiny\color{codegray},
    stringstyle=\color{codepurple},
    basicstyle=\ttfamily\footnotesize,    
    breaklines=true
}
\def\BibTeX{{\rm B\kern-.05em{\sc i\kern-.025em b}\kern-.08em
    T\kern-.1667em\lower.7ex\hbox{E}\kern-.125emX}}

\usepackage{balance}
\usepackage{textcomp}

\def\cuquantum{\textit{cuQuantum}\xspace}
\def\cuquantumsdk{\textit{cuQuantum SDK}\xspace}
\def\cuquantumpy{\textit{cuQuantum Python}\xspace}
\def\cuquantumapp{\textit{cuQuantum Appliance}\xspace}
\def\cusv{\textit{cuStateVec}\xspace}
\def\cutn{\textit{cuTensorNet}\xspace}
\def\cut{\textit{cuTENSOR}\xspace}
\def\path{\textit{path}\xspace}

\newcommand{\ignore}[1]{}

\begin{document}

\title{cuQuantum SDK: A High-Performance Library for Accelerating Quantum Science}

\author{
\IEEEauthorblockN{
Harun Bayraktar\IEEEauthorrefmark{1},
Ali Charara\IEEEauthorrefmark{1},
David Clark\IEEEauthorrefmark{1},
Saul Cohen\IEEEauthorrefmark{1},
Timothy Costa\IEEEauthorrefmark{1},
Yao-Lung L. Fang\IEEEauthorrefmark{1},
}
\IEEEauthorblockN{
Yang Gao\IEEEauthorrefmark{1},
Jack Guan\IEEEauthorrefmark{1},
John Gunnels\IEEEauthorrefmark{1},
Azzam Haidar\IEEEauthorrefmark{1},
Andreas Hehn\IEEEauthorrefmark{1},
Markus Hohnerbach\IEEEauthorrefmark{1},
}
\IEEEauthorblockN{
Matthew Jones\IEEEauthorrefmark{1},
Tom Lubowe\IEEEauthorrefmark{1},
Dmitry Lyakh\IEEEauthorrefmark{1},
Shinya Morino\IEEEauthorrefmark{1},
Paul Springer\IEEEauthorrefmark{1},
Sam Stanwyck\IEEEauthorrefmark{1},
}
\IEEEauthorblockN{
Igor Terentyev\IEEEauthorrefmark{1},
Satya Varadhan\IEEEauthorrefmark{1},
Jonathan Wong\IEEEauthorrefmark{1},
and
Takuma Yamaguchi\IEEEauthorrefmark{1}}
\IEEEauthorblockA{\IEEEauthorrefmark{1}\textit{NVIDIA Corp}, Santa Clara, CA, USA}
}

\maketitle

\begin{abstract}
    We present the NVIDIA \cuquantumsdk~\cite{cuquantum}, a state-of-the-art library of composable primitives for GPU-accelerated quantum circuit simulations. 
As the size of quantum devices continues to increase, making their classical simulation progressively more difficult,
the availability of fast and scalable quantum circuit simulators becomes vital for quantum algorithm developers,
as well as quantum hardware engineers focused on the validation and optimization of quantum devices.
The \cuquantumsdk was created to accelerate and scale up quantum circuit simulators developed by the quantum information science community 
by enabling them to utilize efficient scalable software building blocks optimized for NVIDIA GPU-based platforms.
The functional building blocks provided cover the needs of both state vector- and tensor network- based simulators,
including approximate tensor network simulation methods based on matrix product state,
projected entangled pair state, and other factorized tensor representations.
By leveraging the enormous computing power of the latest NVIDIA GPU architectures,
quantum circuit simulators that have adopted the \cuquantumsdk demonstrate significant acceleration,
compared to CPU-only execution, for both the state vector and tensor network simulation methods.
Furthermore, by utilizing the parallel primitives available in the \cuquantumsdk, one can easily transition to distributed GPU-accelerated platforms,
including those furnished by cloud service providers and high-performance computing systems deployed by supercomputing centers,
extending the scale of possible quantum circuit simulations.
The rich capabilities provided by the \cuquantumsdk are conveniently made available via both Python and C application programming interfaces,
where the former is directly targeting a broad Python quantum community and the latter allows tight integration with simulators written in any programming language.
\end{abstract}

\begin{IEEEkeywords}
quantum circuit simulation, GPU computing, state vector, tensor network
\end{IEEEkeywords}

%%%%%%%%%%%%%%%%%%%%%%%%%%%%%%%%%%%%%%%%%%%%%%%%%%%%%%%%%%%%%%%%%%%%%%%%%%%%
\section{Introduction}\label{sec:intro}
%%%%%%%%%%%%%%%%%%%%%%%%%%%%%%%%%%%%%%%%%%%%%%%%%%%%%%%%%%%%%%%%%%%%%%%%%%%%

\begin{figure}[!hbtp]
    \centering
    \includegraphics[width=.48\textwidth]{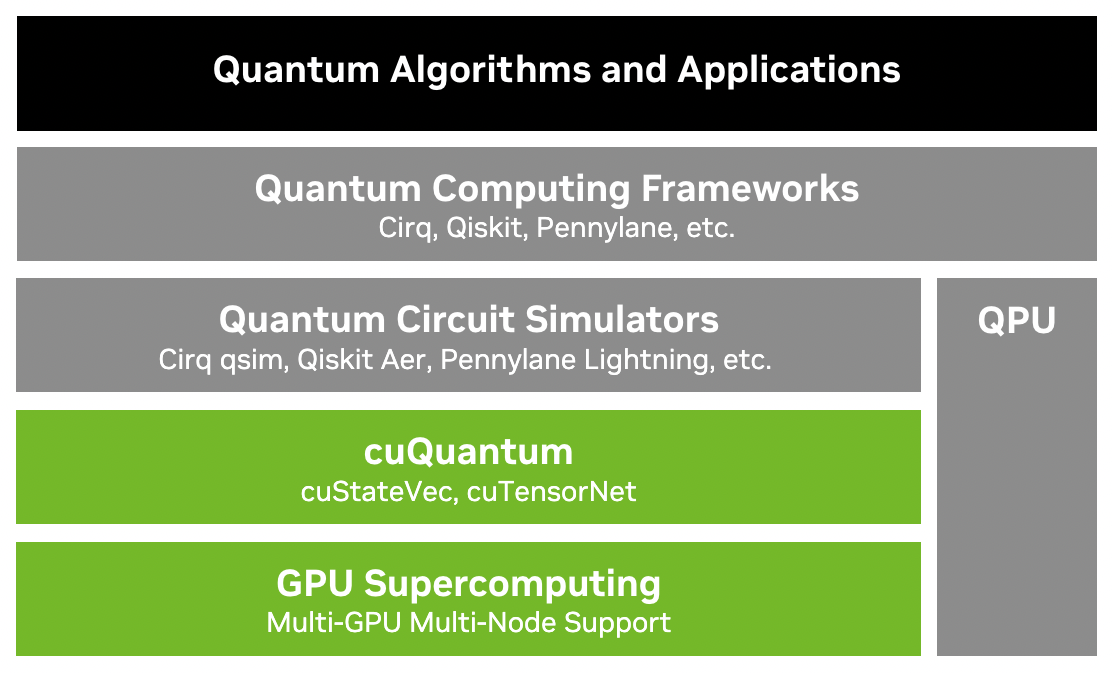}
    \label{fig:cuquantum_overview}
    \caption{Overview of the NVIDIA \cuquantumsdk. QPU stands for quantum processing units.}
\end{figure}

Quantum circuit simulators are a critical part of quantum algorithm and application development workflows.
Today's quantum computers are prohibitively small, error-prone, hard to access, capacity-constrained, and, at times, expensive.
Even as they scale, this will not likely change until Fault-Tolerant Quantum Computing (FTQC) systems
are broadly deployed and no longer capacity-constrained. Therefore, researchers and developers
rely on quantum circuit and analog simulators as a critical tool in their toolbox. Many of these simulators
are based on state vector (SV) and tensor network (TN) simulation methods, both of which rely heavily on
linear algebra and matrix/tensor multiplications. Graphics Processing Units (GPUs) have traditionally been great
computational engines for these types of problems, given their ability to utilize thousands of threads to efficiently
parallelize computations. For these reasons, NVIDIA introduced the \cuquantumsdk with its two main component libraries,
\cusv and \cutn (Fig.~\ref{fig:cuquantum_overview}). Our strategy is focused on accelerating and scaling up all quantum circuit simulators on GPUs.
By working to improve GPU kernels and provide other performance enhancements, in addition to enabling
advanced simulation techniques, we have provided simulator developers around the world with the ability
to perform quantum circuit simulations at scales and speeds previously not available to them.

Before \cuquantum, quantum circuit simulator developers predominantly used basic linear
and tensor algebra libraries in their computational backends. The \cuquantum model has raised
the level of abstraction by providing more convenient and flexible building blocks targeting
quantum circuit simulator developers. At the same time, the provided data structures and
computational primitives are restricted to the core subset of features typically exposed
by higher-level libraries, for example existing TN libraries like
iTensor \cite{iTensor}, Quimb \cite{Quimb}, ExaTN \cite{ExaTN}, Cyclops Tensor Framework \cite{Cyclops},
TensorNetwork \cite{TensorNetworkSoft}, or TensorTrace \cite{TensorTrace}, to name a few.
This uniquely positions \cuquantum as a library created to broadly benefit
the existing quantum information science (QIS) software ecosystem rather than compete
with some of its components. That is, while the primary adoption target is
quantum circuit simulators, \cuquantum can also boost the performance of more generic
software libraries used in quantum sciences.

\cuquantum was introduced to the public in September 2021, with three main components: \cusv, \cutn, and \cuquantumpy.
At the time we introduced support for NVIDIA A100 and V100 devices, followed by a quick expansion of supported systems
to include all devices with CUDA compute capability 7.0+. We demonstrated the largest TN based
simulation of the QAOA MaxCut problem, achieving a 93\% accurate solution to a 10,000 vertex MaxCut graph (5000 qubits).
In March 2022, we introduced the first \cuquantumapp as a Docker container with a Cirq frontend and \cuquantum backend.
The 22.07 release added major performance enhancements for multi-GPU simulations, enabling noisy quantum circuit simulations
at scale in the \cuquantumapp. We also introduced APIs for TN slicing and enabling multi-GPU multi-node execution in \cutn.
In September 2022, we demonstrated the fastest multi-node SV simulations in the world,
exceeding the previous state-of-the-art by over $3x$, and scaling much further. The 22.11 release included
this functionality in the \cuquantumapp, introducing IBM Qiskit as a frontend.
This release of \cutn featured support for the NVIDIA Hopper GPU, user-friendly multi-node APIs,
and approximate TN contraction primitives supporting the matrix product state (MPS),
matrix product operator (MPO), and projected entabled pair state (PEPS) based simulation algorithms.
The 23.03 release added multi-node APIs in \cusv and intermediate caching/reuse in \cutn.
Additionally, we introduced the \cuquantumapp VMI on several major cloud service providers.

The overarching goal of \cuquantum is to fulfill a pressing need from
community QIS codes to accelerate and scale up quantum simulations to
help drive quantum applications towards new discoveries. Our effort
builds upon, focuses on, and unifies decades of code development for
NVIDIA GPU architectures. By adopting \cuquantum, quantum circuit
simulators receive the unique ability to study extremely large quantum
simulations using NVIDIA DGX and HGX systems, cloud based systems,
and some of the largest parallel supercomputers.
The \cuquantumsdk encapsulates SV,
TN, and approximate TN simulation methods which
allow studying and accelerating a wide range of quantum simulations
at unprecedented scale. Our software is the result of the combined efforts
of theoretical physicists, computational scientists, and applied mathematicians
who have formulated effective methods and algorithms, allowing users to explore
quantum algorithms for a wide range of qubit counts and circuit depths.

Despite its relatively short history,
\cuquantum has already demonstrated significant impact by
accelerating the QIS research community worldwide, from industry to academia. Supporting results include the world record for supercomputing scale simulation performance \cite{devblog_abci} and a wide range of novel quantum research, such as optimization \cite{QUARK, Parallelcircuits, Fastcircuitcutting, MakhanovAirbus}, simulating quantum chemistry and other quantum systems \cite{BPDE, Timecrystals}, security and privacy \cite{SupercheQ, QMLwDP}, and much more.

%%%%%%%%%%%%%%%%%%%%%%%%%%%%%%%%%%%%%%%%%%%%%%%%%%%%%%%%%%%%%%%%%%%%%%%%%%%%
\section{The \cusv Library}\label{sec:cusv_lib}
%%%%%%%%%%%%%%%%%%%%%%%%%%%%%%%%%%%%%%%%%%%%%%%%%%%%%%%%%%%%%%%%%%%%%%%%%%%%
We developed the \cusv library to accelerate and scale up SV simulation,
a brute-force, exact method of quantum circuit simulation.
An SV simulator represents the quantum state as a complex-valued 1D vector.
Thus, it behaves as if it were an ideal quantum computer, as the entire quantum state is preserved in the simulator.
Users can successively apply quantum logic gates as needed in their quantum algorithms.
Qubits can be measured without collapsing the quantum state by numerical computation of the probability distribution. 

In order to hold the entire quantum state, the size of a state vector grows exponentially as $2^n$, where $n$ is the number of qubits.
When the size of the state vector exceeds the capacity of a single device, it needs to be distributed across multiple compute devices and nodes connected by high-speed interconnects.
Hence, large-scale SV simulation is considered a high-performance computing (HPC) challenge.

In this section, we first describe several key features of \cusv, followed by an introduction to the SV simulators provided in the NVIDIA \cuquantumapp container.

%Key features
\subsection{\cusv API Design}
The \cusv library was developed to provide a set of common primitives used in SV simulators.
The latest version of the \cusv library, v23.03, provides the features shown in Tab.~\ref{tab_cusv_feature}.
The design of the \cusv library is based on the following considerations: 

\begin{description}[style=unboxed]
   \item[Provide a set of SV primitives with a relaxed memory model:] The \cusv library provides the basic GPU-accelerated APIs such as gate application, measurement, expectation value computation, and sampling. The state vector should be allocated in GPU memory, however, there is no other specific requirement for the memory model. This facilitates the adoption of the \cusv library in existing SV simulators.
   \item[Small memory footprint:] As SV simulation can require a huge amount of memory to store the SV, most APIs are designed to execute in-place operations to eliminate the use of both source and destination buffers.  Most operations are executed by using a small amount of the library-internal temporary buffer attached to the library handle, which therefore does not increase the amount of memory required for simulations.
   \item[Extensible to multi-GPU and multi-node simulations:] Most \cusv APIs work as single GPU primitives. However, these APIs are designed to work as a part of multi-GPU and multi-node simulations. The library provides a set of APIs to swap the index bits of a SV. These are used to manage the index bit ordering of the state vector.  There are multi-GPU and multi-node versions to perform qubit reordering used for distributed SV simulations.
\end{description}

\begin{table}[htbp]
\caption{Key Features of cuStateVec}
\begin{center}
\begin{tabular}{|c|P{6cm}|}

\hline
\textbf{Feature} & {\textbf{Description}} \\ 
\hline
\begin{tabular}{c} Resource \\ management \end{tabular} & \sparbox{6ex}{The library handle holds a few tens of MB of GPU memory and other resources.}\\
\hline
\begin{tabular}{c} Gate \\ application \end{tabular} & \sparbox{6ex}{Gate application with dense matrix and generalized permutation matrix.}\\
\hline
Pauli rotation & \sparbox{3ex}{Rotation of state vector by Pauli string.}\\
\hline
Measurement & \sparbox{8ex}{Measurement on Z-product basis and batched single-qubit measurement. Probability computation and collapse function are separately provided.}\\
\hline
Expectation & \sparbox{6ex}{Expectation value computation with dense matrix observable or Pauli strings.}\\
\hline
Sampler & \sparbox{6ex}{Sampling bit-strings as measurement outcomes without collapsing state vector.}\\
\hline
Accessor & \sparbox{6ex}{Copy state vector between CPU and GPU while manipulating the qubit ordering.}\\
\hline
\begin{tabular}{c} Index \\ bit swap \end{tabular} & \sparbox{6ex}{Swap index bit pairs in the state vector. Single-GPU, multi-GPU, and multi-node versions are available.}\\
\hline

\end{tabular}
\end{center}
\label{tab_cusv_feature}
\end{table}

%Performance Evaluation of Gate Application and Gate Fusion
\subsection{Gate Application Performance and Gate Fusion}\label{sec:cusv_gate_fusion}
\begin{figure}[htbp]
\centerline{\includegraphics[width=.48\textwidth]{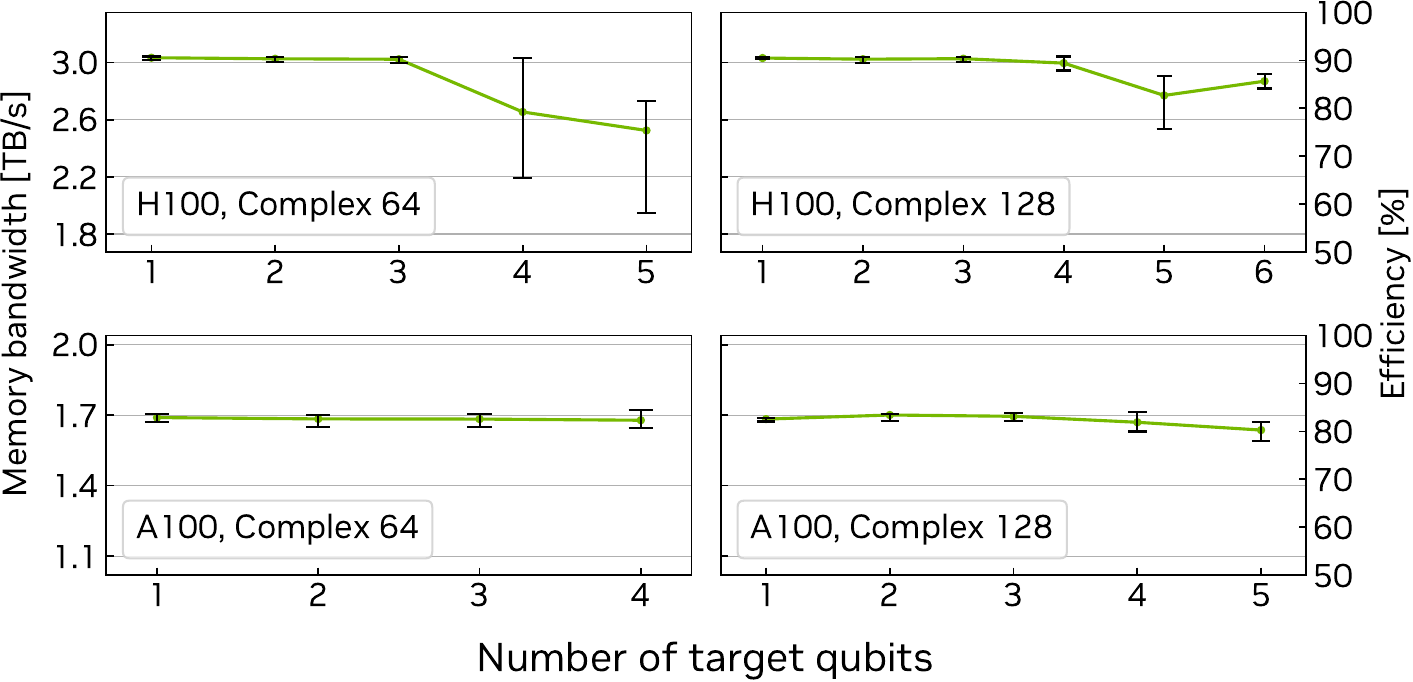}}
\caption{Gate application performance on the NVIDIA H100 80GB SXM and NVIDIA A100 80 GB SXM GPUs. The performance was measured with 33- and 32-qubit state vectors for Complex 64 and Complex 128, respectively. The bandwidth was calculated from 100 measurement results with randomly-selected sets of target qubits.}
\label{fig:cusv_gate_perf}
\end{figure}

In typical SV simulations, gate application -- applying the quantum logic gates from a quantum circuit to a state vector -- is the most time-consuming operation.
While most gates in quantum circuits have just one or two target qubits, \cusv provides optimized gate application APIs for larger numbers of target qubits.
Fig.~\ref{fig:cusv_gate_perf} shows the performance of gate applications on both an NVIDIA H100 80GB SXM GPU and an NVIDIA A100 80 GB SXM GPU, with peak memory bandwidths of 3.35 TB/s and 2.04 TB/s, respectively.  On the H100, the utilized memory bandwidth generally surpasses 2.35 TB/s and reaches 3.0 TB/s for the best cases, corresponding to 70\% and 90\%, respectively, while on the A100, a sustained bandwidth of 1.6 TB/s and a peak of 1.7 TB/s, 78\% and 83\%, respectively, was measured.
For up to 5 qubits for Complex 64 and up to 6 qubits for Complex 128 on an NVIDIA H100 (4 and 5 qubits, respectively on the NVIDIA A100), gate application is a memory-bound operation.
The total simulation time increases in proportion to the number of these gate application API calls as long as the gate matrices are sufficiently small.

To reduce the computation cost, we can introduce gate fusion~\cite{smelyanskiy2016qhipster}.
By fusing numerous small gate matrices into a single, multi-qubit gate matrix, one can apply the fused gate matrix in one shot instead of repeating the application of small gate matrices.
Thus, the number of fused gates directly contributes to the acceleration of the simulations.
Gate fusion can yield drastic performance acceleration in cases where both the high memory bandwidth and high compute performance of the GPU are utilized. 

%Overview of distributed state vector simulation
\subsection{Distributed State Vector Simulation}
\begin{figure}[htbp]
    \centerline{\includegraphics[width=7cm]{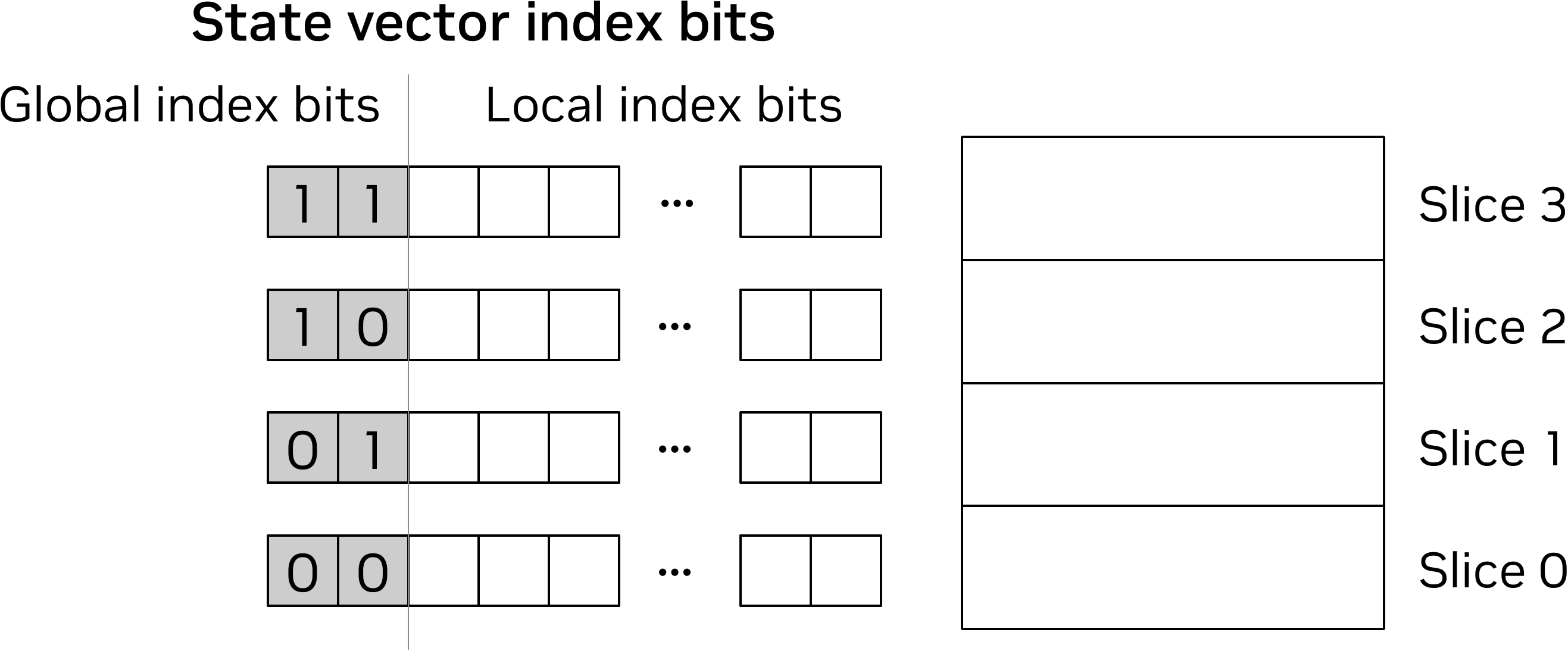}}
    \caption{State vector distribution. The state vector is sliced into four slices and distributed to four computing devices.}
    \label{fig:cusv_sv_distribution}
\end{figure}

In SV simulation, each bit in the state vector index corresponds to one qubit in a quantum circuit.
During simulations, qubits are mapped to the index bits of the state vector.
For distributed SV simulations, the state vector is equally sliced and distributed to multiple computing devices as shown in Fig.~\ref{fig:cusv_sv_distribution}.
In this configuration, the index bits in a slice of the state vector are referred to as local index bits; similarly, the index bits to identify the state vector ordinal are referred to as global index bits.
When applying a gate onto a local index bit, the gate matrix can be applied independently and concurrently to each slice.
However, applying a gate onto a global index bit requires accessing multiple slices of the state vector.

\begin{figure}[htbp]
    \centerline{\includegraphics[width=8.5cm]{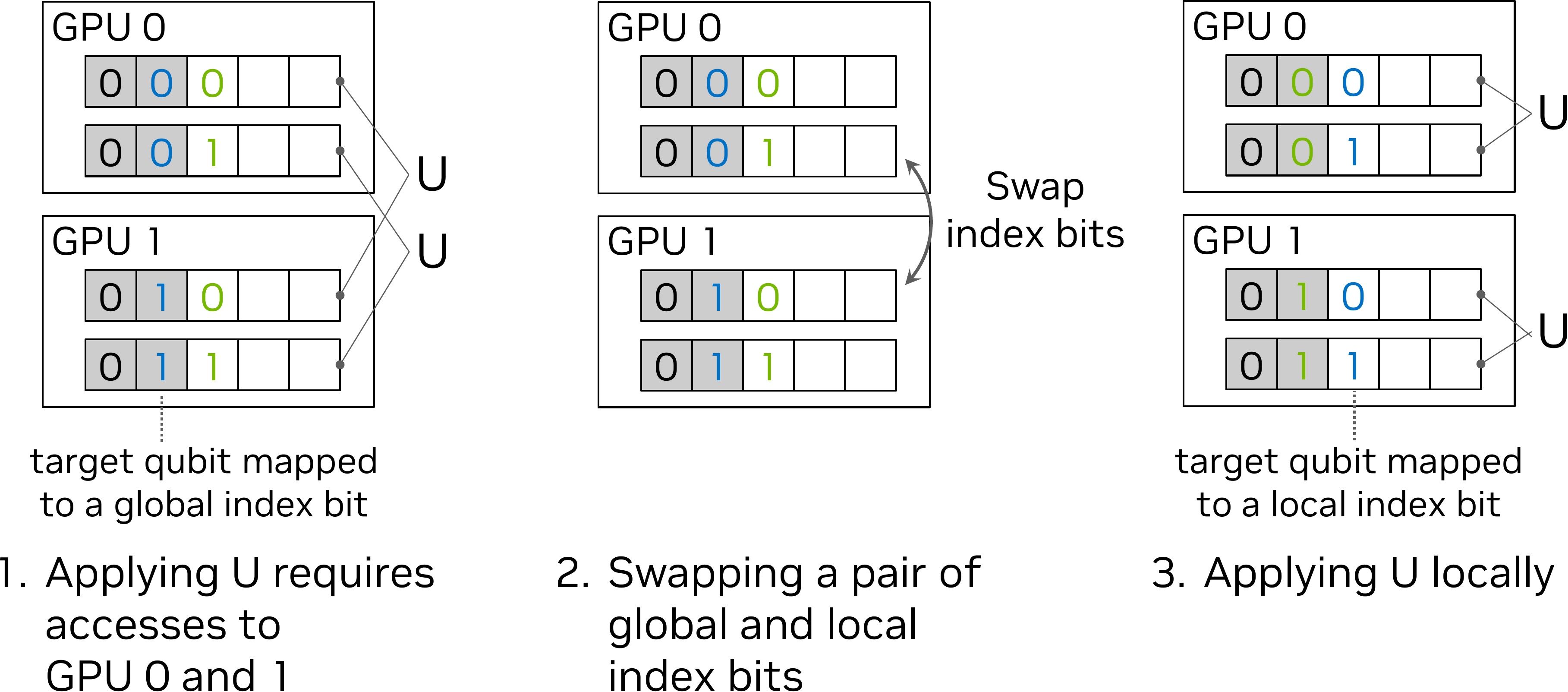}}
    \caption{Qubit reordering by swapping a pair of global and local index bits.  Two state vector slices are distributed to GPU 0 and GPU 1.}
    \label{fig:cusv_qubit_reordering}
\end{figure}

Qubit reordering is a known technique to resolve this limitation, which has been reported in ~\cite{de2007massively}.
As shown in Fig.~\ref{fig:cusv_qubit_reordering}, one can move a global index bit into a local index bit position, and update the qubit-to-index-bit mapping to move a qubit to the local index bits.
Therefore, a gate that acts on a remapped qubit is applied to a local slice of the state vector.
For this purpose, the \cusv library provides distributed index bit swap APIs.
One can develop a distributed SV simulator by keeping target qubits local to the state vector slices, then single-GPU APIs can be applied to update the state vector.
In the \cuquantumapp container, both multi-GPU and multi-node quantum circuit simulations are built with index bit swap APIs and single-GPU \cusv APIs.

%cuQuantum Appliance - qsim/cirq, cusvaer/qiskit-aer simulators
\subsection{Simulators in the Appliance Container}
The NVIDIA \cuquantumapp container provides our multi-GPU-optimized qsim backend for the \textit{Cirq} frontend, and a \textit{cusvaer} backend that supports distributed SV simulations for the \textit{Qiskit Aer}~\cite{qiskit_aer} frontend.
In this section, we provide an overview of each backend.

\subsubsection{Multi-GPU qsim backend}
\textit{qsim}~\cite{isakov2021simulations, qsim2020source} is a Schr\"{o}dinger full SV simulator written in C++ and has been integrated with Cirq~\cite{cirq2021source}, a Python framework for quantum circuit simulation.
Originally, its backends were designed for computations with either multithreading CPUs or single GPUs.
We provide a new backend, \textit{qsim-mgpu}, with multi-GPU functionalities.
We introduce qubit reordering in our backend so that there is no device-to-device communication within each gate application kernel call.
When the reordering is required, we call \texttt{custatevec\-Multi\-Device\-Index\-BitSwaps} to swap the sub state vector components on multiple devices.

\paragraph{Introduction of Dense/Diagonal Gate Fusion}
Some typical gate matrices, e.g., a controlled-Z gate, can be represented as diagonal matrices.
Gate applications for diagonal matrices have lower arithmetic intensity than those for dense matrices due to their sparsity.
Therefore, for diagonal matrices, gate application kernels can exhibit nearly identical performance across a wide range of matrix sizes.

Considering these characteristics, we extend gate fusion in Sec.~\ref{sec:cusv_gate_fusion} to accept diagonal matrices and generate fused diagonal matrices.
In our backend, we provide two options for gate fusion, \texttt{max\_fused\_gate\_size} and \texttt{max\_fused\_diagonal\_gate\_size}, which set the maximum number of qubits per fused dense and diagonal gate, respectively.
The gate applications for dense and diagonal matrices are executed using the \texttt{custatevecApplyMatrix} and \texttt{custatevec\-Apply\-Generalized\-Permutation\-Matrix} APIs.

\paragraph{Performance Evaluations}
In this section, we report the performance of our qsim backend.
We use a single DGX A100 node that consists of eight NVIDIA A100 80GB SXM GPUs and two AMD EPYC 7742 64-core CPUs, a single DGX H100 node that consists of eight NVIDIA H100 80GB SXM GPUs and two Intel Xeon Platinum 8480C CPUs, and version 23.03 of the \cuquantumapp container.
We target Quantum Fourier Transform (QFT)~\cite{weinstein2001implementation}, Quantum Approximate Optimization Algorithm (QAOA)~\cite{farhi2014quantum} with the parameter $p=2$, and Quantum Volume~\cite{cross2019validating} (depth=30) circuits.
Quantum Volume is measured for 10 different circuits and their average elapsed time is evaluated. The maximum gate fusion size is set to 4 on the DGX A100 and 5 on the DGX H100 by the \texttt{max\_fused\_gate\_size} option.
We focus on the 33-qubit Complex 64 problems and measure strong scaling using up to 8 GPUs, using the timer in the qsim backends.

Tab.~\ref{tab_qsimperf} summarizes the elapsed time in each circuit.
Compared to the 1-GPU cases, the 8-GPU cases on the DGX H100 attained 4.6-, 7.0-, and 6.5-times speedups in QFT, QAOA, and Quantum Volume circuits, respectively.
Additionally, we computed the same QFT simulation on CPUs with the \texttt{qsim\_cpu} backend.
The elapsed time was 78.6 s; therefore, our backend with 8 GPUs achieved a 297-fold speedup over the CPU backend.

\begin{table}[htbp]
\caption{\textit{qsim-mgpu} Backend Performance on DGX A100 and DGX H100}
\begin{center}
\begin{tabular}{|c|c|c|c|c|c|c|}
\hline
& & & \multicolumn{2}{|c|}{\textbf{DGX A100}} & \multicolumn{2}{|c|}{\textbf{DGX H100}}\\
\cline{4-7} 
\multirow{2}{*}{\textbf{Circuit}} & \textbf{\# of} & \textbf{\# of} & \textbf{\textit{\# of}} & & \textbf{\textit{\# of}} & \\
& \textbf{GPUs} & \textbf{gates} & \textbf{\textit{fused}} & \textbf{\textit{Time (s)}} & \textbf{\textit{fused}} & \textbf{\textit{Time (s)}} \\
& & & \textbf{\textit{gates}} & & \textbf{\textit{gates}} & \\
\hline
\multirow{4}{*}{QFT} & 1 & 577 & 18 & 2.30 & 18 & 1.21 \\
                     & 2 & 577 & 30 & 1.59 & 27 & 0.911 \\
                     & 4 & 577 & 32 & 0.895 & 29 & 0.523 \\
                     & 8 & 577 & 33 & 0.474 & 27 & 0.265 \\
\hline
\multirow{4}{*}{QAOA} & 1 & 1650 & 131 & 10.7 & 90 & 4.85 \\
                      & 2 & 1650 & 132 & 5.74 & 91 & 2.71 \\
                      & 4 & 1650 & 131 & 2.90 & 88 & 1.34 \\
                      & 8 & 1650 & 132 & 1.48 & 89 & 0.692 \\
\hline
        & 1 & 480 & 154 & 12.6 & 114 & 6.92 \\
Quantum & 2 & 480 & 158 & 6.96 & 122 & 3.91 \\
Volume  & 4 & 480 & 158 & 3.67 & 120 & 2.04 \\
        & 8 & 480 & 160 & 1.93 & 120 & 1.07 \\
\hline
\end{tabular}
\end{center}
\label{tab_qsimperf}
\end{table}

\subsubsection{Qiskit/Qiskit Aer Multi-node Simulator}
Starting with \cuquantumapp v22.11, \textit{cusvaer} is provided for multi-node distributed SV simulations.
\textit{Qiskit Aer}~\cite{qiskit2023source} provides a full-fledged multi-node simulator, enabled at compile time, to \textit{Qiskit} users, and \textit{cusvaer} is an extension to \textit{Qiskit Aer} that enables a new multi-node simulator, optimized to extract the best performance from HPC clusters built with NVIDIA GPUs.

The performance limitation of distributed SV simulations comes from data transfer that happens among distributed slices of the state vector due to necessary qubit reordering.
Using the DGX A100 as a representative GPU server, NVLink and NVSwitch enable high-speed communication between eight GPUs in one node.
NVLink provides 600 GB/s of bidirectional bandwidth for GPU-to-GPU data transfers.
NVSwitch connects eight GPUs in a single node with a bisectional bandwidth of 2.4 TB/s.
In clusters based on the DGX SuperPOD~\cite{dgx_superpod} reference architecture, eight Mellanox ConnectX-6 cards in a DGX A100 are utilized for inter-node communication, which provides 200 GB/s of unidirectional bandwidth to/from one DGX A100 node.
The \textit{cusvaer} backend is designed and optimized to achieve nearly-optimal performance, accelerating distributed SV simulations.

\begin{figure}[htbp]
    \centerline{\includegraphics[width=8cm]{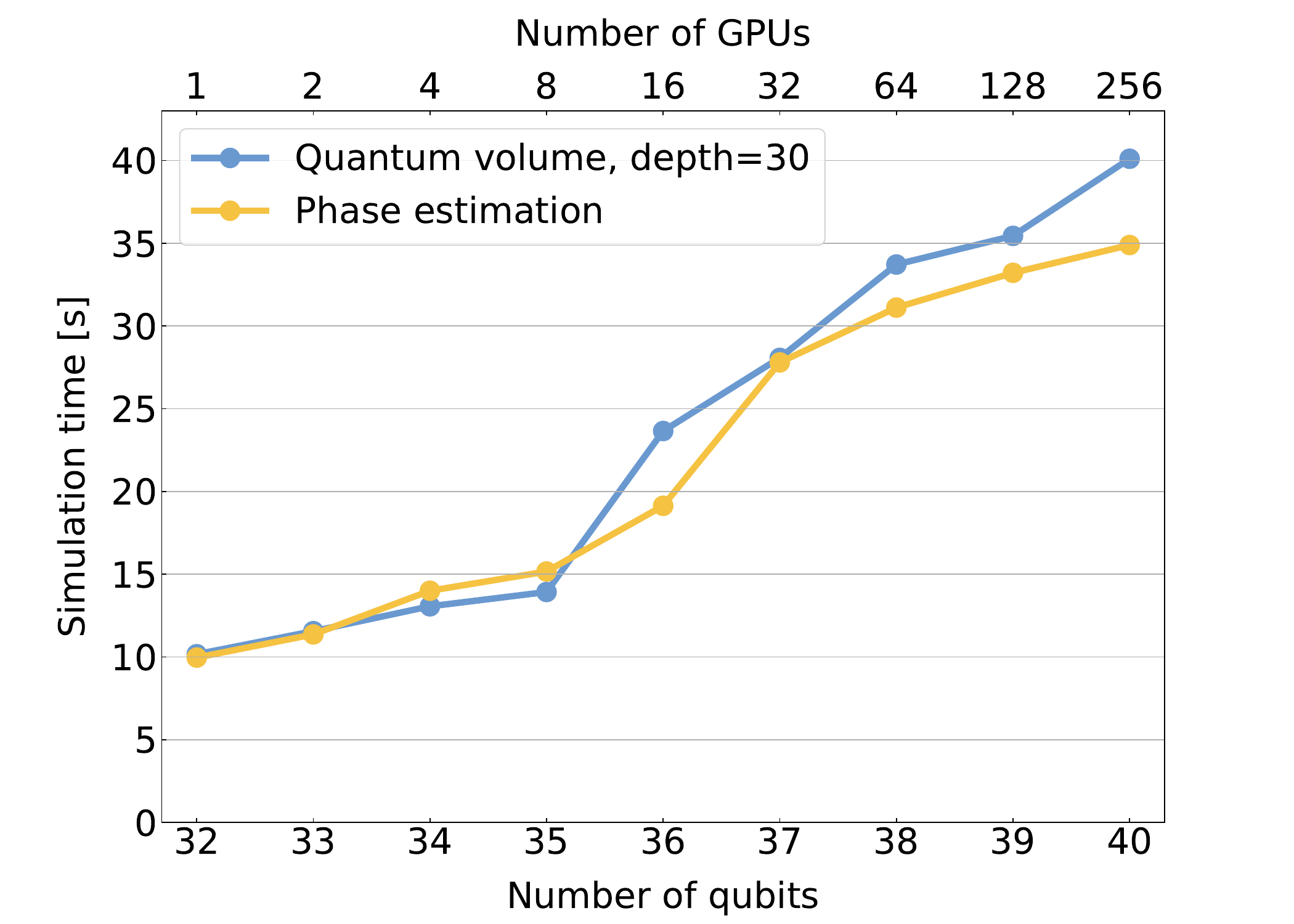}}
    \caption{The simulation time of the extended \textit{Qiskit Aer} multi-node simulator on the Selene supercomputer.}
    \label{fig:cusv_cusvaer}
\end{figure}

Fig.~\ref{fig:cusv_cusvaer} shows the performance results of the multi-node SV simulator for Quantum Volume (depth=30) and quantum phase estimation~\cite{cleve1998quantum}.
The performance has been measured on NVIDIA's Selene supercomputer, which is an NVIDIA-internal DGX A100 cluster based on the DGX SuperPOD architecture.
The number of qubits is varied from 32 qubits (1 GPU, 1 node) to 40 qubits (256 GPUs, 32 nodes) for Complex 128 value type.
From 32 to 35 qubits, the number of GPUs doubled in a single node, where NVLink and NVSwitch are utilized for the data transfer for qubit reordering.
From 35 qubits, the number of nodes is doubled for each addition of one qubit.
The slope of the simulation time from 35 to 40 qubits is steeper than that from 32 to 35 qubits, which reflects the difference in data transfer bandwidths between NVLink/NVSwitch and IB network.
These results suggest that it is necessary to implement algorithms to reduce inter-node and inter-device communications such as qubit reordering for better scalability.
The simulation times of 32 qubit circuits with two sockets of AMD EPYC 7742 64-core CPU were 178 seconds for quantum volume (depth=30) and 102 seconds for QFT, while the simulation times for 40 qubit circuits of our multi-node simulator are less than 40 seconds.  The performance shown in Fig.~\ref{fig:cusv_cusvaer} is one of the best among multi-node SV simulators.
The performance comparison with other SV simulators is discussed in the NVIDIA Developer Blog~\cite{devblog_abci}.

%%%%%%%%%%%%%%%%%%%%%%%%%%%%%%%%%%%%%%%%%%%%%%%%%%%%%%%%%%%%%%%%%%%%%%%%%%%%
\section{The \cutn Library}\label{sec:cutnlib}
%%%%%%%%%%%%%%%%%%%%%%%%%%%%%%%%%%%%%%%%%%%%%%%%%%%%%%%%%%%%%%%%%%%%%%%%%%%%

Tensor network theory provides powerful and versatile numerical machinery that
has been successfully employed across multiple quantum domains, including condensed
matter physics \cite{OrusTN, schollwock2011density}, quantum chemistry \cite{ChanDMRG}, and
quantum information science (QIS) \cite{markov2008simulating, GrayTN, zhou2020limits},
as well as applications in other domains such as probabilistic graphical
models \cite{ZhangMPScalculus} and machine learning \cite{StoudenmireML, EvenblyML}.
In all cases, TNs enable the exploration and exploitation of low-rank structure
in inherently multi-dimensional problems such as quantum many-body simulations.
TN methods have found numerous applications in quantum
circuit simulations. The most straightforward approach is based on the direct conversion of a quantum
circuit to the corresponding TN, followed by its contraction \cite{markov2008simulating}.
While quite powerful for low-depth high-qubit-count quantum circuits \cite{VillalongaRQCSummit,
GrayTN, AlibabaRQC, NVIDIAQAOA}, this approach will
suffer from exponential scaling once the circuit depth grows sufficiently large.
More sophisticated TN methods tend to introduce controllable approximations
with tolerable errors by enforcing an approximate representation of the quantum
circuit wave-function or density matrix. For example, methods based on the
MPS \cite{zhou2020limits, StoudenmireQFT, StoudenmireGrover,
LyakhMPS} or PEPS \cite{guo2019general} form of compression.
These approximate techniques are extremely powerful for simulating quantum circuits
with a moderate degree of quantum entanglement \cite{MPSChemSunway}.

Given that the classical simulation of quantum phenomena is often highly amenable to GPU acceleration,
there was a clear need for a library of performant generic building blocks which would make feasible
an efficient implementation of the TN methods inside the corresponding domain-specific
libraries and simulators. The \cutn library consists of two main modules.
The network contraction module that performs contraction of a TN and the approximate tensor module
that performs exact or approximate tensor(s) decomposition. 
The \cutn library from the \cuquantumsdk offers
both C and Python APIs (see Sec.~\ref{sec:cuqnt_python}). The APIs are flexible,
exposing most of the features implemented in the library allowing users to control,
explore, and investigate the minor details of the algorithmic techniques.

%part a cutn overview
%%%%%%%%%%%%%%%%%%%%%%%%%%%%%%%%%%%%%%%%%%%%%%%%%%%%%%%%%%%%%%%%%%%%%%%%%%%%
\subsection{Tensor Network Contraction Module}\label{subsec:cutn_contraction}
%%%%%%%%%%%%%%%%%%%%%%%%%%%%%%%%%%%%%%%%%%%%%%%%%%%%%%%%%%%%%%%%%%%%%%%%%%%%
TN contractions are computed as a sequence of pairwise tensor
contractions (called a \path). 
Defining which pair of tensors go together and the order of the
pairwise contractions are the most crucial and complex phase in TN
contraction. 
The ratio between the cost (computational and memory) of an optimal or {\it close-to-optimal} 
and a naive path (random, left to right, right to left) 
can differ by many orders of magnitude emphasizing the 
importance of the \path.
Once the \path is defined, there is another phase of
optimization related to computing the pairwise contractions using the most efficient GPU
kernels. As a result, the TN contraction module of \cutn can be described as a
combination of two components, a pathfinder component that runs on the CPU and an 
execution component that computes contractions on the GPU.

%---------------------------------------------------------------------------
\subsubsection{Path Finding}\label{subsubsec:pathfinding}
%---------------------------------------------------------------------------
The role of a pathfinder is to find a contraction path that minimizes the cost
of contracting the TN. Finding an optimal contraction path is an NP-hard problem; 
its cost grows exponentially with the size of the network, making
such a technique an impractical or unrealistic solution even for small networks of
size dozens of tensors. Most TN simulations, in particular,
quantum circuit simulations, consist of networks of hundreds of tensors.
Different techniques such as optimal, greedy, or branching~\cite{opt_einsum}
have been developed, but they either provide a far from
optimal path or require exponential time to find a path.
The approach taken in \cutn is similar to the one presented in~\cite{GrayTN},
where it was shown to provide superior quality paths ({\it close-to-optimal}).
We developed many algorithmic advancements and optimization techniques
to quickly provide such high-quality paths.
It starts by simplifying the network. Simplification is
a technique that preprocesses the large TN to find all sets of
straightforward contractions. It removes them from the network and
replaces each set with its final tensor. The result is a smaller tensor network that
is easier to process. \cutn then uses a hyper-optimization technique~\cite{GrayTN} that
tunes the pathfinder procedure. The core of the pathfinder engine is based on
recursive graph partitioning combined with a ``bubbling" mechanism. The
bubbling method can be described as an iterative process that selects hot spots in
the contraction tree (e.g., subtrees that are the most expensive in terms of
flops or memory) and adjusts their cost using multiple pathfinder algorithms
suitable for small networks. The hyper-optimizer loop
creates a distinct set of configurations of parameters for the pathfinder
engine and picks the best path found. This hierarchical and recursive design
embedded into the hyper-optimizer loop provides superior quality contraction paths (see Fig.~\ref{fig:cutn_path_quality}).
Higher quality contraction paths allow larger simulations to be tackled,
opening new frontiers in research and discovery.

%---------------------------------------------------------------------------
\subsubsection{Slicing}\label{subsubsec:slicing}
\begin{figure}[!hbtp]
    \centering
    \includegraphics[width=.45\textwidth]{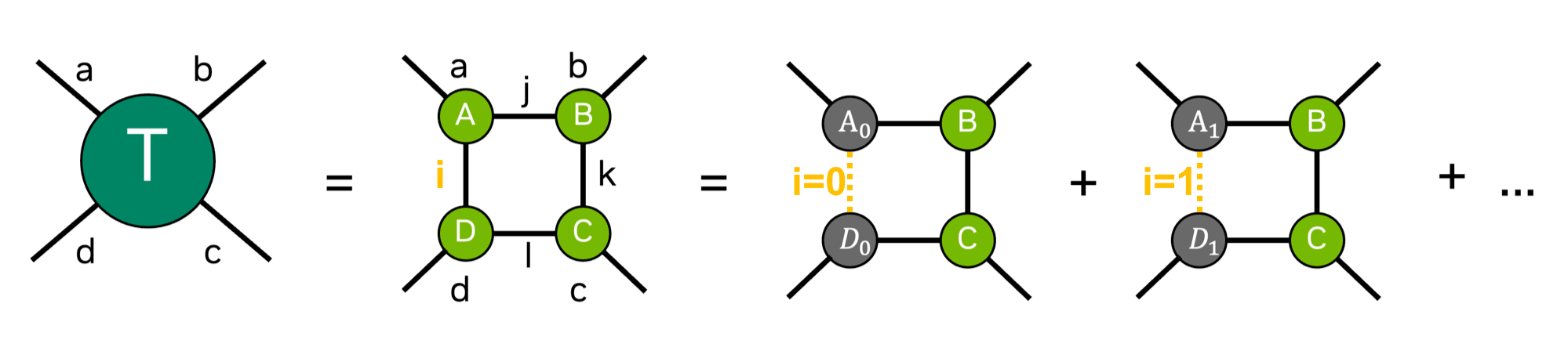}
    \caption{Overview of the slicing technique.}
    \label{fig:cutn_slicing}
\end{figure}

%---------------------------------------------------------------------------
Slicing is a technique implemented in \cutn to either make a network
contraction fit into available device memory or to create parallelism for
distributed execution. Large TN contractions need more memory than
that available in a single device. Techniques like out-of-device memory might be
implemented in such situations but at the cost of a large performance drop due
to the significant data traffic between the device and the CPU memory required in this case. Realistic
simulations, or the type of simulations needed to develop new horizons in quantum
computing, may require more memory than is available on GPU devices or on CPU
making it impossible to perform without a different approach. By slicing (also known as
variable projection or bond cutting), we split the contraction of the entire
TN into several independent smaller contractions, where each
contraction considers only one particular position of a certain mode (or
combination of modes). The result for full TN contraction can be
obtained by summing over the output of each sliced contraction. In
Fig.~\ref{fig:cutn_slicing}, we illustrate an example. If we slice over the mode {\it i},  it
is no longer implicitly summed over as part of the tensor contraction, but instead
explicitly summed in the last step. As a result, slicing can effectively reduce
the memory footprint for contracting large TNs, particularly
those arising from quantum circuits.

Since each sliced contraction is independent of the others, the
computation can be efficiently parallelized in various distributed settings. As
a result, the slicing techniques can also be used to generate parallelism and to
speed up TN contractions even if memory is not an issue. 

Despite all the benefits above, the downside of slicing is that it often
increases the total FLOP count of the entire contraction. The overhead of
slicing heavily depends on the contraction path and the modes that are sliced.
In general, there is no straightforward way to determine the best set of modes
to slice. To increase the probability of finding the best contraction path and
the best slicing plan, we integrated this phase inside the pathfinder module
that is itself encapsulated inside the hyper-optimizer. 

%---------------------------------------------------------------------------
\subsubsection{Planning and Workspace}\label{subsubsec:planning}
%---------------------------------------------------------------------------
\cutn utilizes the NVIDIA \cut library~\cite{cutensor} as a backend to perform all the pairwise tensor contractions in the TN, 
over the existing GPU devices.
Once a contraction path has been generated (or received from the user) for the TN, 
a contraction plan needs to be constructed, This will hold the pairwise contraction plans for \cut.
Such a contraction plan could be used to contract the network multiple times, possibly with different data each time.
Given the multiple ways a network can be contracted, in terms of the order of contracting the tensor pairs within the same contraction path, 
the contraction plan is by default configured to use the contraction order that utilizes the minimum workspace possible during network contraction.
However, since the provided workspace memory has a direct impact on the choice and performance of the underlying \cut kernels, 
\cutn provides APIs to query the minimum workspace memory size required, the recommended size for good performance, 
and the maximum size the contraction plan can utilize.
The user can then supply any workspace size larger than the minimum required.
To get the best performance when contracting the network, \cutn offers an API to automatically tune the contraction plan, 
by executing multiple \cut kernels on each of the pairwise contractions and selecting the most performant one 
that can operate within the scratch workspace-size constraint, 
as well as optimizing the intermediate tensor shapes to best utilize device throughput and minimize data transfers.

%---------------------------------------------------------------------------
\subsubsection{Contraction}\label{subsubsec:contraction}
%---------------------------------------------------------------------------
\cutn facilitates the contraction of the whole TN using the contraction plan, 
while leveraging the high performance of \cut on GPU devices.
The contraction plan serves as the vehicle for repeatedly contracting the TN,
each time with possibly different data while using the same contraction plan.
The contraction API also facilitates the contraction of the network per slice. 
It is possible to select the whole set of slices to contract at once, 
or select a set of slices, picked by indices or by ranges, to be contracted at a contraction call,
with the ability to accumulate results on the target buffer or overwrite it.
This flexibility allows for integration of the contraction process with user workflows 
and easy distribution of the contraction computation to many hardware resources,
as will be discussed in a subsequent section.

%---------------------------------------------------------------------------
\subsubsection{Performance}\label{subsubsec:performance}
\begin{figure}[!hbtp]
    \centering
    \includegraphics[width=.45\textwidth]{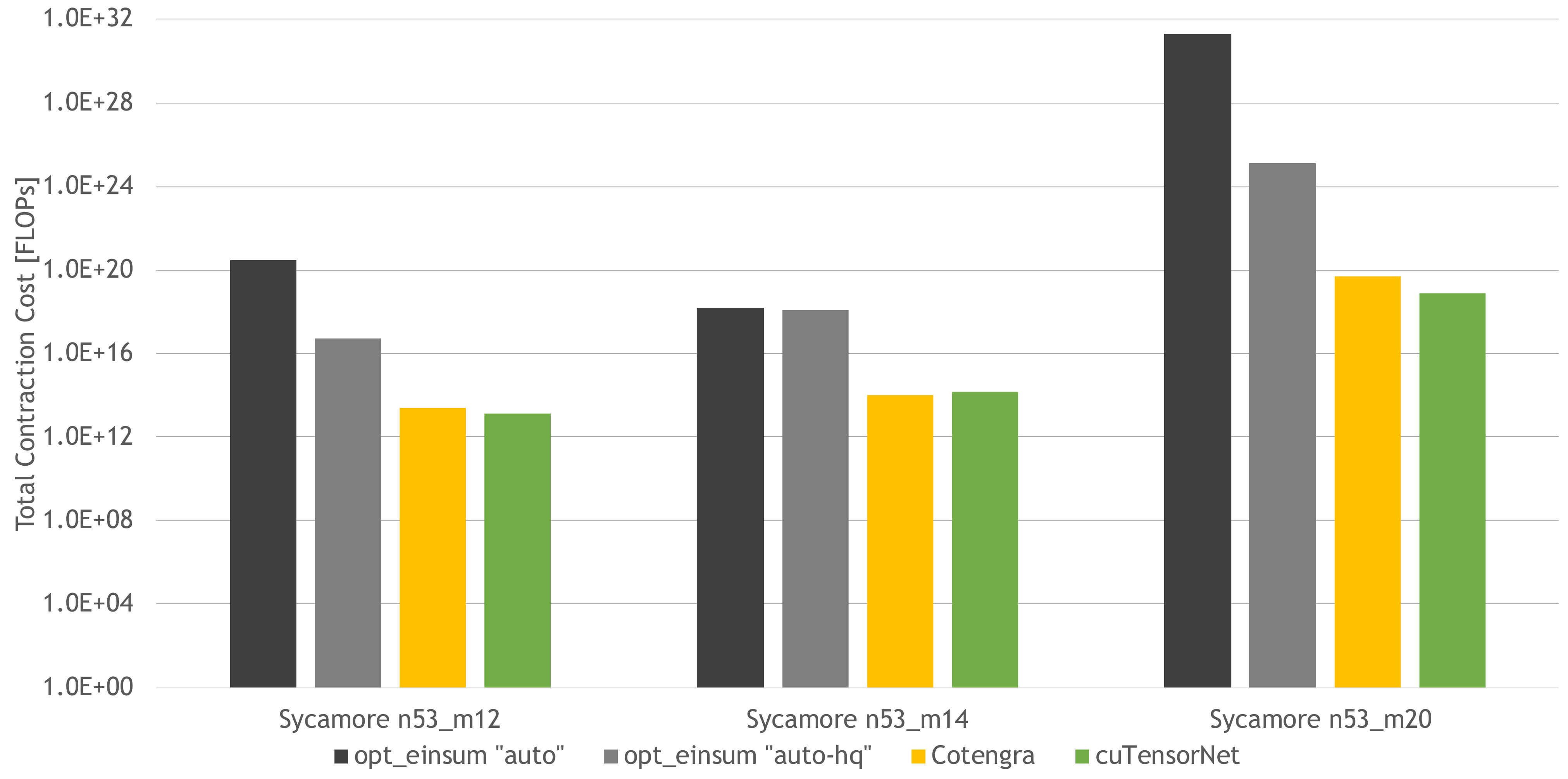}
    \caption{\cutn pathfinder quality of the resulting contraction path compared to similar packages, measured in FLOPs.}
    \label{fig:cutn_path_quality}
\end{figure}

%---------------------------------------------------------------------------
There are two relevant metrics when considering the performance of a
pathfinder: the quality of the path found, and the time taken to find that
path. The former is plotted in Fig.~\ref{fig:cutn_path_quality}, measured by the cost of the
obtained contraction path in FLOPS. The circuits used for benchmarking are the
random quantum circuits at depths 12, 14, and 20 described
in~\cite{GrayTN}. \cutn performs significantly better than the optimized
opt\_einsum library~\cite{opt_einsum} in finding a {\it close-to-optimal} path
and slightly better than Cotengra~\cite{GrayTN} for these circuits. 

\cutn also finds a high-quality path quickly. The time in seconds required 
by \cutn compared to Cotengra~\cite{GrayTN} to run 1000 hyper-optimizer samples
is depicted in 
Tab.~\ref{tab:cutn_path_time},
for Sycamore-53 quantum circuits with different depths. For the most
complex problem, with over 3,000 tensors in the network, \cutn takes an average of 
about 8 seconds per path compared to 730 seconds for Cotengra.

\begin{table}[htbp]
    \caption{Pathfinding time (sec), for the Sycamore RQC problems on Intel(R) Core(TM) i9-7900X CPU.}
    \begin{center}
        \begin{tabular}{|l||c|c|c|}
        \hline
            {Time in seconds for}      & \textbf{n53\_m10} & \textbf{n53\_m20} & \textbf{n53\_m20 } \\
            {1000 hyper-opt samples}      & \textbf{168 tensors} & \textbf{ 382 tensors} & \textbf{3316 tensors} \\
        \hline
            Cotengra~\cite{GrayTN}    & $1.5~10^3$  & $8.8~10^4$    & $7.3~10^5$ \\
            \cutn                       & $~~2~10^2$   & $1.4~10^3$   & $~~8~10^3$ \\
        \hline
        \end{tabular}
    \end{center}
\label{tab:cutn_path_time}
\end{table}
 
For realistic quantum simulation, the time to find a path is considered to be
negligible when compared to the cost of performing the contraction of the
network. \cutn performs the contraction on GPUs using the \cut library that
provides tuned and optimized kernels for each GPU architecture. \cutn also
implements internal optimizations to further reduce the memory footprint required
to perform all the pairwise contractions, as well as a mechanism to reorder the intermediate tensors'
modes to realize the best performance of the GPU kernels. 

\begin{figure}[!hbtp]
    \centering
    \includegraphics[width=.45\textwidth]{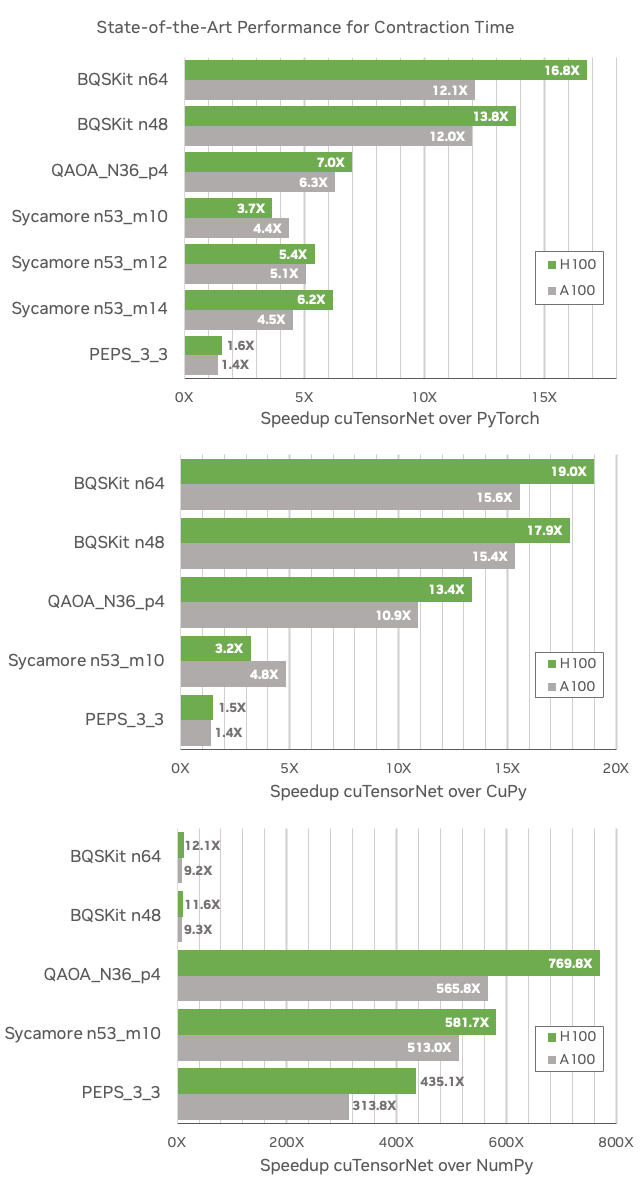}    
    \caption{Contraction speedup of \cutn vs PyTorch~\cite{pytorch} and
    CuPy~\cite{cupy} running on a single NVIDIA H100 80GB SXM GPU or NVIDIA A100 80GB SXM GPU, as well as
    vs NumPy~\cite{numpy} running on an AMD EPYC 7742 8-core CPU for
    several key quantum benchmarks.} 
    \label{fig:cutn_execution_time}
\end{figure}

Fig.~\ref{fig:cutn_execution_time} depicts the speedup of the network
contraction using \cutn compared to PyTorch~\cite{pytorch} and CuPy~\cite{cupy}
running either on a single NVIDIA H100 80GB SXM GPU or a single NVIDIA A100 80GB SXM GPU.
Fig.~\ref{fig:cutn_execution_time} also illustrates the speedup of \cutn compared to
NumPy~\cite{numpy} running on AMD EPYC 7742 8-core CPUs with multithreaded OpenBLAS,
for several key quantum 
benchmarks ranging from search-based circuit synthesis to quantum approximate optimization algorithm
to Sycamore random quantum circuits.
Depending on the circuit, \cutn offers 
as much as a 19x speedup for the execution of the network contraction when compared to
PyTorch and CuPy.
\cutn shows impressive speedup when compared to NumPy running multi-threaded on the CPU.

%part b distributed
\subsection{Distributed Multi-GPU Multi-Node Execution}\label{subsec:cutn_mgmn}
\begin{figure}[h]
    \centering
    \includegraphics[width=.45\textwidth]{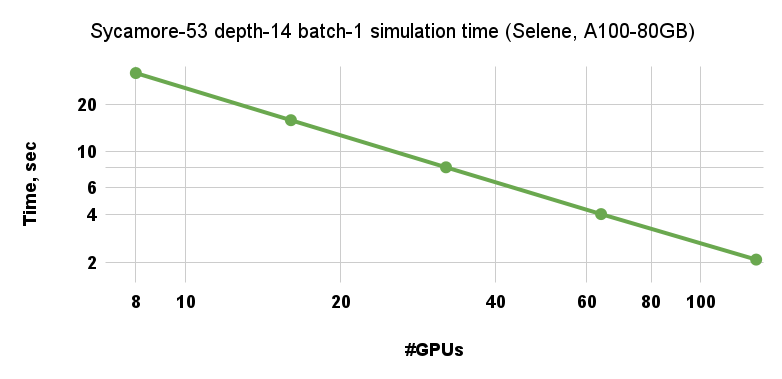}
    \caption{Strong scaling of the Sycamore-53 random quantum circuit (RQC) simulation (circuit depth = 14)
             in which a single bit string probability amplitude is computed.}
    \label{fig:mgmn_benchmark}
\end{figure}

In order to enable and further accelerate larger quantum circuit simulations,
\cutn introduced automatic multi-GPU multi-node (distributed)
parallelization in the 22.11 release.
By activating distributed parallelization, quantum applications which have already adopted \cutn
for single-GPU acceleration can immediately transition to large-scale GPU-accelerated cloud
platforms and HPC systems by leveraging one of the Message Passing Interface (MPI) libraries,
thus scaling up the size of possible quantum simulations.
The activation of distributed parallelization is performed via a single
API call to the \texttt{cutensornetDistributedResetConfiguration} function
which takes a user-provided MPI communicator. Once activated,
both the \textit{pathfinder} and the \textit{execution} procedures
will be parallelized across all MPI processes associated with the given MPI communicator.
The \textit{pathfinder} will distribute hyper-sampling of contraction path candidates
and the execution procedure will distribute the generated tensor network slices.
Since both procedures are embarrassingly parallel, one can expect strong
scaling to a very large number of GPUs. Fig.~\ref{fig:mgmn_benchmark} shows
a strong scaling plot for a simulation of a single bit string probability amplitude
of a random quantum circuit that was part of the validation experiments
on Google's Sycamore quantum chip in 2019 \cite{QuantumSupremacy}. The simulation was run
on the Selene supercomputer at NVIDIA, with up to 128 Ampere A100-80GB GPUs.
The simulation time scales almost ideally with the number of GPUs.

Since different MPI libraries do not necessarily conform to the same ABI specification,
special care had to be taken in interfacing \cutn with MPI. In particular,
the same \cutn library will work with any standard-conforming MPI implementation
that supports CUDA. This is achieved by dynamically loading a thin shared library
(at run-time) built from the provided cuTensorNet-MPI interface C source code
and linked to user's MPI library.

% reuse part c
\subsection{Intermediate Tensor Caching}\label{subsec:cutn_reuse}

\subsubsection{Intermediate Tensor Reuse}\label{subsubsec:reuse_intro}
In quantum circuit simulations, some tasks may require multiple evaluations of TNs
of the same structure where only a small subset of input tensors undergoes a value update
in repeated executions. In particular, validation of quantum processors often requires computing
probability amplitudes of individual bit-strings consistently\cite{QuantumSupremacy}. In this case, the tensor
networks for all requested bit-strings have the same structure, only differing in the
values of the output bits. Another relevant example is the quantum circuit sampling procedure
which requires repetitive computation of projected reduced density matrices (RDMs). In this case, one needs
to recompute the same TN (projected RDM) many times where only a small
subset of input tensors change their values (projected output bits).
The simulation of these and similar cases can be accelerated by storing and reusing the intermediate tensors
which stay constant across all repeated TN evaluations.
To mark input tensors as constant or mutable,
\textit{cuTensorNet} provides the corresponding API.
To facilitate reuse of intermediate tensors,
\textit{cuTensorNet} accepts two kinds of workspace memory, 
\textit{scratch} (used to hold temporary data during computations) 
and \textit{cache} (used to store the constant intermediate
tensors upon the first network contraction call to enable their reuse in subsequent TN
contraction calls). 

\subsubsection{Performance Impact}\label{subsubsec:reuse_perf}
Intermediate tensor reuse through caching can bring drastic speed-ups to the contraction of TNs
where only a small subset of input tensors (mutable tensors) change their value across many network contractions.
Tab.~\ref{tab_reuseperf} shows the huge performance impact of intermediate tensor reuse, on a synthetic network,
where the number of input tensors is indicated as network size, while varying the number of constant input tensors, 
and running 1000 repetitions of TN contraction.

\begin{table}[htbp]
    \caption{Intermediate tensor reuse performance impact on NVIDIA H100 80GB SXM GPU and NVIDIA A100 80GB SXM GPU with 1000 repetitions of TN contraction}
    \begin{center}
        \newlength{\width}
        \width.1\linewidth
        \begin{tabular}{|p{\width}|p{\width}|p{\width}|p{\width}|p{\width}|p{.23\linewidth}|}
        \hline
        \textbf{Network Size} & \textbf{\# slices} & \textbf{Constant tensors} & \textbf{Speedup H100} & \textbf{Speedup A100} & \textbf{Utilized / Recommended Cache (GB)} \\
        \hline
        \multirow{3}{*}{242} & \multirow{3}{*}{1} & 80\% & 4.19 X & 4.12 X & 1.21 / 1.21 \\
                                                & & 85\% & 4.25 X & 4.22 X & 1.10 / 1.10 \\
                                                & & 90\% & 4.32 X & 4.24 X & 1.11 / 1.11 \\
        \hline
        \multirow{3}{*}{327} & \multirow{3}{*}{1024} & 80\% & 1.92 X   & 1.71 X   & 55.34 / 1099.51 \\
                                                   & & 85\% & 815  X   & 804 X & 4.29 / 4.29 \\
                                                   & & 90\% & 838  X   & 827 X & 4.29 / 4.29 \\
        \hline
        \end{tabular}
    \end{center}
\label{tab_reuseperf}
\end{table}

%approxTN part d
\subsection{Approximate Tensor Network Features}\label{subsec:cutn_approx}

\subsubsection{Overview}\label{subsubsec:approx_overview}
At the core of prevalent approximate TN methods are numerical techniques, such as QR and Singular Value Decomposition (SVD), that can be employed to efficiently represent the quantum state. 
For instance, QR decomposition is heavily used in MPS canonicalization for orthogonalization, while SVD is employed to truncate the bond dimension, a parameter that determines the accuracy of the MPS \cite{mcculloch2007density, schollwock2011density}. 
While these two decomposition techniques are well-defined at the matrix level, extending them to the tensor level can be described as a transpose-decompose-transpose-transpose process,
similar to the way tensor contraction is generalized from matrix multiplication. 
In practice, tensor decomposition is frequently used in conjunction with contractions to compress the network graph in a controlled manner. 

\subsubsection{Functionalities}\label{subsubsec:approx_features}

\cutn provides APIs that target different levels of single and compound tensor operations in a hierarchical manner, 
aiming to support the needs of the numerous approximate TN algorithms. 
At the tensor level, \cutn provides C APIs for both QR and SVD functionalities on a single GPU device, leveraging fast cuSOLVER kernels~\cite{cusolver} for transposition and decomposition.
For the tensor SVD API, truncation of singular values and the corresponding U and V tensors, as well as post-processing of output tensors, are supported in addition to the standard exact decomposition.

\begin{figure}[h]
    \centering
    \includegraphics[width=.45\textwidth]{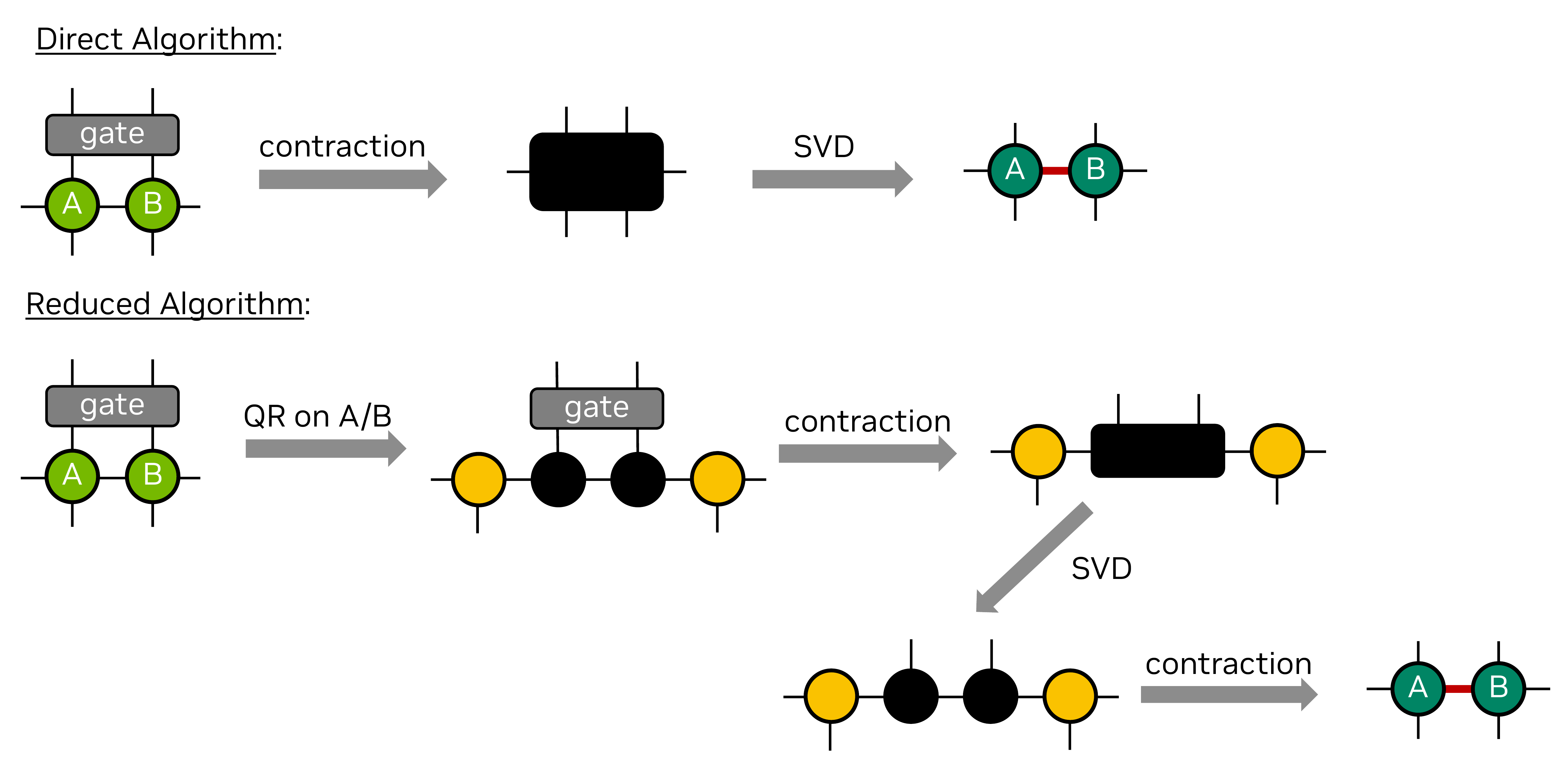}
    \caption{Two algorithms for the gate split operation. The upper panel shows the direct algorithm, while the lower panel presents the reduced algorithm.}
    \label{fig:gate_split}
\end{figure}

At the TN level, \cutn provides a specialized C API named \texttt{cutensornetGateSplit} that targets a compound operation frequently used in constructing a TN representation of a quantum circuit, such as MPS.
In this context, the gate split process refers to a computational task wherein the gate operand is factorized onto two connecting tensors that it acts upon.
Fig.~\ref{fig:gate_split} illustrates two algorithms supported in \cutn for performing the gate split task.
The direct algorithm involves a full contraction followed by decomposition, while the reduced algorithm uses QR decomposition before the full contraction to potentially reduce the intermediate size for the SVD computation.

\subsubsection{Performance}\label{subsubsec:approx_performance}

\begin{figure}[h]
    \centering
    \includegraphics[width=.48\textwidth]{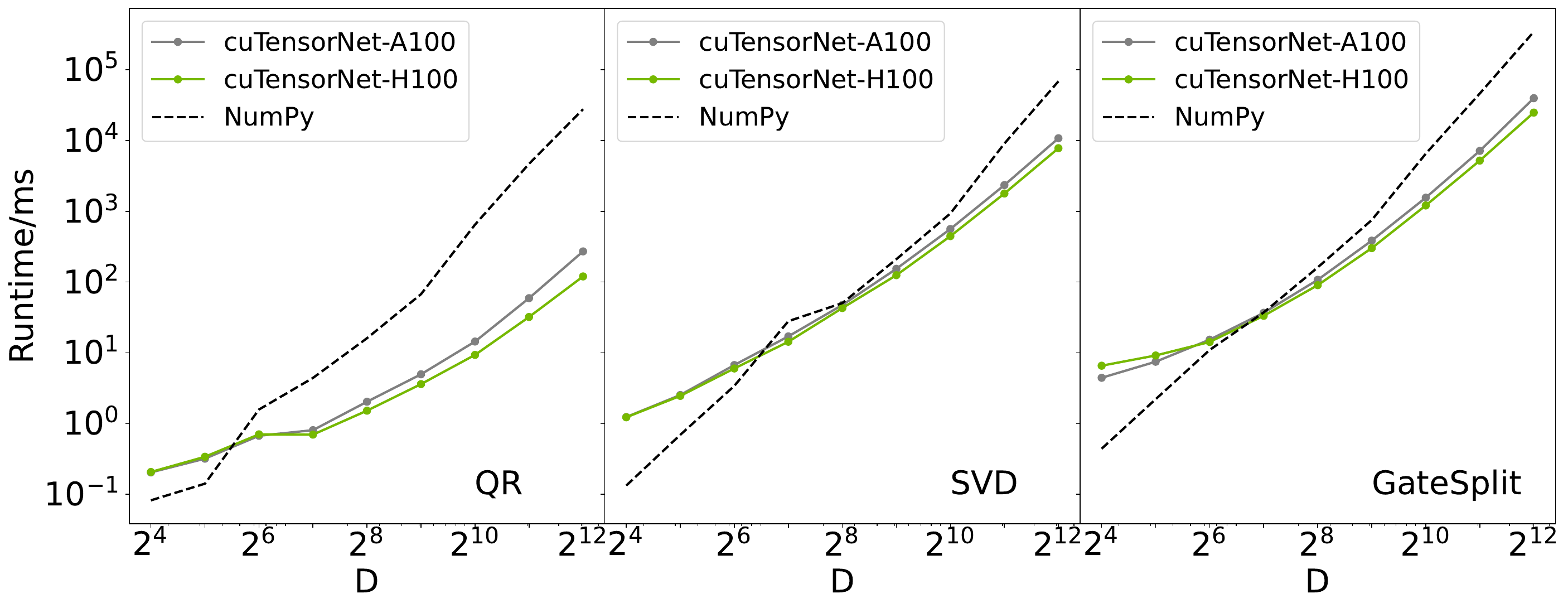}
    \caption{Performance benchmark of tensor QR, SVD and GateSplit. The execution time for the NumPy implementation is indicated by the dashed black line, while that of the cuTensorNet APIs is shown by the solid green lines.}
    \label{fig:approx_benchmark}
\end{figure}

In this subsection we present performance benchmarks for the C APIs, including tensor QR, SVD, and \textit{GateSplit}, on NVIDIA A100 and H100 80GB GPUs. 
For the QR and SVD benchmarks, we measure the the execution time of decomposing a rank-3 MPS tensor with a shape of $(D, 2, D)$ as a function of $D$, where $D$ denotes the bond dimension of the MPS.
For the gate split benchmarks, the execution time of \textit{GateSplit} was measured for applying a two qubit-gate to an MPS with a bond dimension $D$.
The performance of \cutn was benchmarked against an equivalent CPU-based (\textit{NumPy}) implementation using all 64 cores of an AMD EPYC 7742 CPU 
and the results are summarized in Fig.~\ref{fig:approx_benchmark}. 
With tensor QR, \cutn exhibits a speedup over the CPU implementation for bond dimensions greater than 32, with a peak of $102$x for A100 and $230$x for H100 at $D=4096$.
The performance curves of tensor SVD and \textit{GateSplit} are similar to that of tensor QR, but speedup is observed at a larger bond dimension. 
The peak speedup of tensor SVD at $D=4096$ reaches $6.4$x for A100 and $8.8$x for H100. Meanwhile, the speedup of \textit{GateSplit} is observed to be $8.6$x and $13.8$x for A100 and H100, respectively.
The similarity between the peak speedup of tensor SVD and \textit{GateSplit} is consistent with the fact that the cost of SVD is expected to dominate the contraction cost at large scales.

\section{cuQuantum Python}\label{sec:cuqnt_python}

One of the goals of the NVIDIA \cuquantumsdk is to allow users to easily access its full functionalities from within Python, enabling interoperability with other Python frameworks and projects.  So, in addition to the C APIs provided in the SDK, we also offer \cuquantumpy as a natural starting point for those who aim to accelerate their Python workloads using NVIDIA GPUs.
To do so, we adopt a two-layer approach:
\begin{enumerate}
\item Provide 1:1 Python bindings for the C APIs of both \cusv and \cutn libraries from the NVIDIA \cuquantumsdk.
\item Provide high-level, pythonic APIs for easier integration with Python applications in QC or other domains.
\end{enumerate}
Generally, we follow PEP 8 as our style guideline, for both 1:1 bindings and pure Python functions. This allows us to offer APIs that feel natural to Python users.
We offer both wheels and conda packages
for users to quickly spin up a working Python environment without compiling anything from source.
Below we briefly introduce each layer.

\subsection{Low-Level Python bindings}\label{sec:cuqnt_py_low}

The low-level bindings are 1:1 mappings of the C public APIs to Python. They are exposed under the library modules \texttt{cuquantum.custatevec} and \texttt{cuquantum.cutensornet}.
They currently are written in Cython~\cite{cython} to support different kinds of user-provided arguments 
so as to offer flexibility, allowing users to explore the trade-offs between performance and convenience brought by Python's dynamic nature.
If a C API requires an array of plain-old-data (POD) structs, for example, we map them to Python using NumPy's structured dtype
~\cite{numpy_datatype_ref}
so that users can allocate such structs using the familiar NumPy APIs, e.g., \texttt{arr = np.zeros((10,), dtype=my{\_}struct{\_}dtype)},
and pass the array to the corresponding Python binding.
The low-level bindings are then used by high-level APIs when possible, as discussed below.

\subsection{High-Level Pythonic APIs}\label{sec:cutn_python}

The high-level pythonic APIs provide encapsulated functionalities that are natural and handy for Python users.  With these APIs, just a few lines of Python code can provide functionality that would otherwise require hundreds of lines of C code.
Our pythonic APIs keep all the low-level boilerplate code, including memory and resource management, away from the concerns of users.
Taking \cutn as example, we offer
\texttt{cuquantum.einsum()} that performs TN contraction following the equivalent
\texttt{numpy.einsum()} API signature and semantics
~\cite{numpy_einsum_semantics_ref}
and, thus, can be used as a drop-in replacement,
\texttt{cuquantum.contract()} that extends \texttt{einsum()} to allow customization and controllability over the tensor
network path finding and contraction execution,
\texttt{cuquantum.\-cutensornet.\-tensor.decompose()} that supports single-tensor decomposition with QR and SVD,
and  \texttt{cuquantum.\-cutensornet.\-experimental.\-contract{\_}decompose()} that provides access to additional functionalities beyond gate split operations, 
enabling it to handle contraction and decomposition operations for arbitrary TNs. 

However, if users need finer control over resource management or prefer less automation, they can also use the underlying
Python classes and methods. For example, the \texttt{cuquantum.Network} class offers the aforementioned encapsulation
of a TN. Its constructor parses an einsum expression with tensor operands to create a TN topology and prepare metadata.
It also has
a \texttt{contract{\_}path} method for finding an optimal path, 
an \texttt{autotune} method for auto-selecting the best contraction kernel,
and a \texttt{contract} method for actual contraction execution.
All of the configurability offered at the C API level is exposed through \texttt{Network\-Options},  \texttt{Optimizer\-Options},
 \texttt{Path\-Finder\-Options}, \texttt{Reconfig\-Options}, and \texttt{Slicer\-Options}.

Moreover, as mentioned above, we can comfortably interact with other Python libraries and frameworks at this level. For example,
the \cutn distributed contraction support can be easily turned on, by passing the MPI communicator from
\textit{mpi4py}~\cite{mpi4py} (a \texttt{mpi4py.MPI.Comm} object) 
to our helper function and then to
\texttt{cuquantum.\-cutensornet.\-distributed{\_}reset{\_}configuration()}, allowing us to support users of differnt
underlying MPI vendors thanks to mpi4py's abstraction.
Another example involves the \texttt{cuquantum.BaseCUDAMemoryManager} protocol for permitting
\cuquantumpy to share and use the same memory allocation solution (e.g. CuPy or PyTorch mempool) from other components of the user's workload,
so as to avoid encountering the out-of-memory issues commonly seen when multiple Python GPU libraries are involved.
Finally, besides using CuPy as the default tensor framework, \cuquantumpy also supports NumPy ndarrays and PyTorch tensors as input tensor operands.

The layered hierachy also allows us to build QC capability upon the lower-level,
``QC-agnostic'' functionalities. For example,
\texttt{cuquantum.CircuitToEinsum} is a Python class specifically designed to parse a user-provided, fully parameterized quantum circuit
(from either the Cirq or Qiskit framework)
and turn it into a TN contraction task, by generating einsum-compatible contraction inputs (with the chosen tensor framework).
The class supports various computation targets, such as SV coefficients, single or batched bit-string amplitudes, RDMs, and expectation values for Pauli strings. 
By specifying the computation target, users can obtain the corresponding contraction inputs that can be
passed to \texttt{einsum()}, \texttt{contract()}, or \texttt{Network()},
using either default settings for expedited computation or a customized path optimization solution. 
Users can optionally enable the so-called reverse lightcone simplification technique \cite{GrayTN} when computing RDMs and expectation values,
so as to reduce the effective TN size and further accelerate the computation.

%%%%%%%%%%%%%%%%%%%%%%%%%%%%%%%%%%%%%%%%%%%%%%%%%%%%%%%%%%%%%%%%%%%%%%%%%%%%
\section{Conclusions}\label{sec:concl}
%%%%%%%%%%%%%%%%%%%%%%%%%%%%%%%%%%%%%%%%%%%%%%%%%%%%%%%%%%%%%%%%%%%%%%%%%%%%

As we have shown, NVIDIA \cuquantumsdk provides flexible and highly optimized
software building blocks for quantum circuit simulator developers and other domain
scientists interested in the efficient GPU implementation of quantum-inspired algorithms.
In particular, the quantum circuit simulation primitives provided by the cuStateVec
library ensure the optimal memory footprint for state vector simulators. The gate
application procedure achieves high memory bandwidth.
State vector simulators which adopted the cuStateVec library have demonstrated
large GPU speed-ups with respect to CPU-only execution. The distributed execution
primitives have been shown to scale well on multi-node multi-GPU platforms, enabling
even larger state vector simulations.

The cuTensorNet library from \cuquantumsdk has been shown to generate high-quality
tensor network contraction paths much faster than the current
state-of-the-art software packages targeting tensor network simulator developers.
The use of the cuTENSOR library as the computational backend, combined with
additional optimizations of tensor network contraction planning, deliver
significant GPU speed-ups in the tensor contraction phase. The intermediate
caching/reuse offers further speed-ups in the workloads based on repetitive
tensor network contractions. The automated distributed multi-node multi-GPU
parallelization provided by the cuTensorNet library enables straightforward
transition of the tensor network simulators which adopted cuTensorNet to
cloud platforms and HPC systems, often showing close-to-ideal scalability.
A set of flexible and performant tensor contraction-decomposition primitives
facilitates efficient implementation of a wide variety of approximate tensor
network contraction schemes.

Finally, to allow easy access to all functionalities of \cuquantumsdk
from within Python frameworks and projects, we developed \cuquantumpy APIs
as a natural extension for those who aim to directly accelerate their Python
workloads using NVIDIA GPUs.

We envisage to continue our progress towards accelerating and scaling up
a more broad circle of quantum applications where we intend to extend our
coverage by developing more functionalities and features essential
for the ongoing and future quantum research efforts.

%%%%%%%%%%%%%%%%%%%%%%%%%%%%%%%%%%%%%%%%%%%%%%%%%%%%%%%%%%%%%%%%%%%%%%%%%%%%
\section*{Acknowledgment}
%%%%%%%%%%%%%%%%%%%%%%%%%%%%%%%%%%%%%%%%%%%%%%%%%%%%%%%%%%%%%%%%%%%%%%%%%%%%

We gratefully acknowledge
Anima Anandkumar, Jean-Marie Eichner, RayLynn Harwood,
Jean Kossaifi, Taylor Patti, Jeremy Wang,
Demy Xu, Huawei Li, Marvel Zeng,
Micky Abir, Marc G.\ Davis, Omid Khosravani, Fereshte Mozafari, and Shi-Ning Sun
for their support and help throughout many phases and releases of the cuQuantum project.
We also acknowledge the
cuTENSOR~\cite{cutensor},
cuSOLVER~\cite{cusolver},
and CuPy~\cite{cupy} teams
for fruitful collaboration,
and the conda-forge community~\cite{conda_forge} and the Kitmaker team for packaging support.

%%%%%%%%%%%%%%%%%%%%%%%%%%%%%%%%%%%%%%%%%%%%%%%%%%%%%%%%%%%%%%%%%%%%%%%%%%%%
%\section*{References}
%%%%%%%%%%%%%%%%%%%%%%%%%%%%%%%%%%%%%%%%%%%%%%%%%%%%%%%%%%%%%%%%%%%%%%%%%%%%
\balance
\bibliographystyle{IEEEtranN}   % for using with natbib
\bibliography{ourbib}
\end{document}